\documentstyle[12pt]{article}
\begin{document}

\author{Mario Castagnino \\
Instituto de Astronom\'{\i}a y F\'{\i}sica del Espacio,\\
Casilla de Correos 67, Sucursal 28,\\
1428 Buenos Aires, Argentina. \and Roberto Laura \\
Departamento de F\'{\i}sica.\\
Facultad de Ciencias Exactas, Ingenier\'{\i}a y Agrimensura,\\
Universidad Nacional de Rosario\\
Av. Pellegrini 250, 2000 Rosario, Argentina.}
\title{Minimal Irreversible Quantum Mechanics:\\
Pure States Formalism}
\maketitle

\begin{abstract}
It is demonstrated that, making minimal changes in ordinary quantum
mechanics, a reasonable irreversible quantum mechanics can be obtained. This
theory has a more general spectral decompositions, with eigenvectors
corresponding to unstable states that vanish when $t\rightarrow \infty .$
These ''Gamov vectors'' have zero norm, in such a way that the norm and the
energy of the physical states remain constant. The evolution operator has no
inverse, showing that we are really dealing with a time-asymmetric theory.
Using Friedrichs model reasonable physical results are obtained, e. g. : the
remaining of an unstable decaying state reappears, in the continuous
spectrum of the model, with its primitive energy.

$\bullet $PACS Nrs. 05.20-y, 03.65. BZ, 05.30-d.

$\bullet $e-mail: castagni@iafe.uba.ar
\end{abstract}

\section{Introduction.}

Usually rigorous quantum mechanics must be formulated in a Gel'fand triplet 
\cite{Bogo}: 
\begin{equation}
{\cal S\subset H\subset S^{\times }},  \label{1.1}
\end{equation}
where:

${\cal S}$ is the space of ''regular states'' or test functions space, that
correspond to Schwarz class wave functions, that are considered as the
physical states.

${\cal H}$ is the space of ''states'', or Hilbert space, introduced to
extend the notion of probability to a larger space. These states are square
integrable wave functions, e. g. Schwarz functions where a finite set of
points is removed from the curve representing the function. As it is not
clear for us what is the physical meaning of this kind vectors${\cal ,}$ in
this paper we will consider that only ${\cal S}$ contains the physical
states.

${\cal S^{\times }\ }$is the space of ''generalized states'', or rigged
Hilbert space, namely the space of linear functionals over ${\cal S},{\cal \ 
}$that essentially are used to find the spectral expansion of the regular
states.

Let $K$ be the Wigner or time-inversion operator. In the usual
time-symmetric or reversible quantum mechanics the evolution Hamiltonian $H$
is time symmetric, i. e.: 
\begin{equation}
KHK^{\dagger }=H.  \label{1.3}
\end{equation}

In fact, if it were time-asymmetric the theory would be trivially time
asymmetric, and we know that such a trivial theory do not coincide with
physical reality. In wave function representation $K$ coincide with complex
conjugation, so it is defined over ${\cal S}$ by: 
\begin{equation}
K\varphi (x)=\varphi ^{*}(x),  \label{1.4a}
\end{equation}
\begin{equation}
K:{\cal S\rightarrow S}.  \label{1.4b}
\end{equation}

So ${\cal S}$ space is also time-symmetric.

But the real universe and macroscopic objects have clearly time-asymmetric
evolutions, so we must explain how these time-asymmetry appears if the
quantum mechanical basic laws of the universe (embodied in ${\cal H}$ )are
time-symmetric. The usual and successful explanation is based in coarse
grainig: macroscopic objects have a huge number of dynamical variables and
we can measure and control only a small number of them, the so call relevant
variables. If we neglect the rest of the variables, the irrelevant ones, we
obtain time-asymmetric evolution equations \cite{20}.

Nevertheless in this paper (according to the line of thought pioneered in
references \cite{Sudar}\cite{Bohm}\cite{anto}) we want to follow a different
way, because we believe that the development of an alternative theory will
enhance our knowledge about time-asymmetry. Thus we want to sketch an
irreversible quantum theory, which explains time-asymmetry from the basic
microscopic level directly. In this way we will have two (probably
equivalent) theories to compare.

Obviously we want to obtain our new theory making minimal changes to the
well established and usual quantum mechanics. If we change eqs. \ref{1.3} or 
\ref{1.4a} it is almost sure to find experimental problems. So the minimal
modification is to change eq. \ref{1.4b} defining a new test functions space 
$\phi _{-}\subset {\cal S}$ such that: 
\begin{equation}
K:\phi _{-}\rightarrow \phi _{+}\neq \phi _{-}.  \label{1.5}
\end{equation}
In this way $K$ is not even defined over the space of regular states $\phi
_{-}$ and time-asymmetry naturally appears.

We shall demonstrate that an irreversible quantum theory based in a Gel'fand
triplet: 
\begin{equation}
\phi _{-}\subset {\cal H}\subset \phi _{-}^{\times },  \label{1.6}
\end{equation}
is feasible and it yield reasonable physical results, as the decaying of
unstable states, if test function space $\phi _{-}$ is properly chosen. We
shall show that, what it is done in the quoted papers \cite{Sudar}\cite{Bohm}%
\cite{anto}, is essentially our minimal modification of the ordinary
reversible quantum theory. But with this new approach we gain a more clear
comprehension of the extension from the reversible quantum theory to the
irreversible one, described in these papers.

The paper is organized as follows:

-In section 2 we review the analytic extension method used to obtain new
spectral decompositions: the main new tool of the formalism. Complex
eigenvalues appear in this expansion corresponding to unstable states or
Gamov vectors.

-In section 3 it is proved that the norm of these Gamov vectors vanish,
showing that they are not physical states but only ''ghosts'', that we can
use to make simpler our computations. This fact is essential to preserve the
norm and the energy of the physical states.

-In section 4 we prove that, in our theory, the time evolution operator has
no inverse. This fact shows that this theory is really time asymmetric. We
can then discuss the origin of the arrow of time.

-In section 5 the Friedrichs model is introduced and some reasonable
physical results are obtained.

In this paper we only deal with pure states. The study of mixed states is in
progress and will be the subject of another paper \cite{Laura}.

\section{Generalized spectral decomposition by analytic extensions.}

\subsection{Usual spectral decomposition of $H_0.$}

Let us consider first ${\cal S}$ as the space of regular wave vectors in
energy representation.

The internal product of two wave vectors $\varphi $ and $\psi $, represented
by $\varphi (\omega )$ and $\psi (\omega )$, both belonging to ${\cal S}$,
is given by 
\begin{equation}
\langle \varphi |\psi \rangle =\int_0^\infty d\omega \varphi ^{*}(\omega
)\psi (\omega )
\end{equation}

Let us suppose that the free Hamiltonian operator $H_0$ satisfy 
\begin{equation}
\langle \varphi |H_0\,\psi \rangle =\langle H_0\,\varphi |\psi \rangle
=\int_0^\infty d\omega \varphi ^{*}(\omega )\,\omega \,\psi (\omega )
\end{equation}

If we now introduce the linear (antilinear) functionals $\langle \omega |$ ($%
|\omega \rangle $) on ${\cal S}$, defined by 
\begin{equation}
\langle \omega |\psi \rangle \equiv \psi (\omega ),\qquad \langle \varphi
|\omega \rangle \equiv \varphi ^{*}(\omega ),
\end{equation}
expressions 2.1 and 2.2 can be written as 
\begin{eqnarray}
\langle \varphi |\psi \rangle &=&\int_0^\infty d\omega \langle \varphi
|\omega \rangle \langle \omega |\psi \rangle , \\
\langle \varphi |H_0\,\psi \rangle &=&\int_0^\infty d\omega \langle \varphi
|\omega \rangle \omega \langle \omega |\psi \rangle
\end{eqnarray}

If we omit the 'bra' $\langle \varphi |$ ('ket' $|\psi \rangle $) in 2.4 we
obtain the following formal expression for $|\psi \rangle $ ($\langle
\varphi |$) in terms of the functionals $|\omega \rangle $ ($\langle \omega
| $): 
\begin{equation}
|\psi \rangle =\int_0^\infty d\omega |\omega \rangle \langle \omega |\psi
\rangle ,\quad \langle \varphi |=\int_0^\infty d\omega \langle \varphi
|\omega \rangle \langle \omega |
\end{equation}

Equations 2.6 yield equation 2.4 for the product $\langle \varphi |\psi
\rangle $, if we impose on the functionals defined in eq. 2.3 the
generalized orthogonality condition 
\begin{equation}
\langle \omega |\omega ^{\prime }\rangle =\delta (\omega -\omega ^{\prime }).
\end{equation}

From 2.4 and 2.5 we can also obtain the formal expressions 
\begin{eqnarray}
I &=&\int_0^\infty d\omega |\omega \rangle \langle \omega | \\
H_0 &=&\int_0^\infty d\omega |\omega \rangle \omega \langle \omega |
\end{eqnarray}

Equation 2.9 is the usual spectral decomposition of $H_0$, satisfying 
\begin{eqnarray}
\langle \omega |H_0\,\psi \rangle &=&\omega \langle \omega |\psi \rangle
\qquad \psi \in {\cal S}  \nonumber \\
\langle \varphi |H_0\,\omega \rangle &\equiv &\langle H_0\varphi |\omega
\rangle =\omega \langle \varphi |\omega \rangle \qquad \varphi \in {\cal S}
\end{eqnarray}

\subsection{Complex spectral decomposition of $H_{0.}$}

In order to obtain more general spectral expansions than the usual ones we
would like to promote the real variable $\omega $ to a complex variable $z$
and to change the integral over ${\cal R}{\bf ^{+}}$ in equation 2.4 by an
integral over a curve $\Gamma $ of the complex plane, as in figure 1. This
change can be done if $|\psi \rangle \in \Psi \subset {\cal S\subset H}$ and 
$|\varphi \rangle \in \Phi \subset {\cal S\subset H}$, where $\Psi $ and $%
\Phi $ are subspaces of ${\cal S\subset H}$, for which the analytic
extensions $\psi (z)$ of $\psi (\omega )$ and $\varphi ^{\#}(z)$ of $\varphi
^{*}(\omega )$ are well defined at least in the shadowed region of fig, 1.
We have introduced the notation $\varphi ^{\#}(z)=[\varphi (z^{*})]^{*}$.
Thus, we can write 
\begin{equation}
\langle \varphi |\psi \rangle =\int_\Gamma dz\,\varphi ^{\#}(z)\,\psi
(z),\qquad \varphi \in \Phi ,\quad \psi \in \Psi
\end{equation}

We define the linear (antilinear) functionals $\langle z|$ $(|z\rangle )$ on 
$\Psi $ $(\Phi )$,by 
\begin{eqnarray}
\langle z|\psi \rangle &\equiv &\psi (z),\qquad \psi \in \Psi  \nonumber \\
\langle \varphi |z\rangle &\equiv &\varphi ^{\#}(z),\qquad \varphi \in \Phi
\end{eqnarray}
and therefore 
\begin{eqnarray}
\langle \varphi |\psi \rangle &=&\int_\Gamma dz\langle \varphi |z\rangle
\langle z|\psi \rangle \\
\langle \varphi |H_0\psi \rangle &=&\int_\Gamma dz\langle \varphi |z\rangle
z\langle z|\psi \rangle
\end{eqnarray}

We can also write the formal expressions 
\begin{eqnarray}
|\psi \rangle &=&\int_\Gamma dz\,|z\rangle \langle z|\psi \rangle  \nonumber
\\
\langle \varphi | &=&\int_\Gamma dz\langle \varphi |z\rangle \langle z|
\end{eqnarray}

From 2.15 we obtain 2.13 if we impose the generalized orthogonality
condition 
\begin{equation}
\langle z|z^{\prime }\rangle =\delta _\Gamma (z-z^{\prime }),
\end{equation}
where $\delta _\Gamma $ is defined by the equation 
\[
\int_\Gamma dzg(z)\delta _\Gamma (z-z)=g(z), 
\]
being $g(z)$ an adequate test function defined on $\Gamma $.

From equations 2.13 and 2.14 we obtain the formal expressions 
\begin{eqnarray}
I &=&\int_\Gamma dz\,|z\rangle \langle z| \\
H_0 &=&\int_\Gamma dz\,|z\rangle \,z\,\langle z|
\end{eqnarray}

Equation 2.18 is the new spectral decomposition for $H_0.$

By the analytic extension of equations 2.10 we obtain 
\begin{eqnarray}
\langle z|H_0\,\psi \rangle &=&z\,\langle z|\psi \rangle \qquad \psi \in \Psi
\nonumber \\
\langle \varphi |H_0\,z\rangle &\equiv &\langle H_0\,\varphi |z\rangle
=z\,\langle \varphi |z\rangle \qquad \varphi \in \Phi
\end{eqnarray}

Thus, $|z\rangle $ and $\langle z|$ are generalized right and left
eigenvectors of $H_0$ ($z$ belong to the shadowed region of figure 1).%
\footnote{%
In the usual spectral decomposition of $H_0$ we have, according to 2.3:
\par
\[
\langle \varphi |\omega \rangle =[\varphi (\omega )]^{*}=[\langle \omega
|\varphi \rangle ]^{*}, 
\]
and therefore the adjoint generalized state of $|\omega \rangle $ is $%
\langle \omega |$.
\par
This is not the case for the complex spectral decomposition. According to
our definition 2.12
\par
\[
\langle \varphi |z\rangle =\varphi ^{\#}(z)\equiv [\varphi
(z^{*})]^{*}=[\langle z^{*}|\varphi \rangle ]^{*}, 
\]
and the adjoint generalized state of $|z\rangle $ is, in this case, $\langle
z^{*}|$.}.So formally: 
\[
\langle z|H_0=z\,\langle z|\qquad H_0\,|z\rangle =z\,|z\rangle 
\]

\subsection{Usual spectral decomposition of H.}

Up to now, we have obtained analytical continuations starting from the
complete set $\{|\omega \rangle \}$ of generalized eigenvectors of the
unperturbed Hamiltonian $H_{0.}$ We can, as well, start from the generalized
eigenvectors $\{|\omega _{\pm }\rangle \}$ of the total Hamiltonian $%
H=H_0+V, $ given by the solutions of the Lipmann-Schwinger equation: 
\begin{eqnarray}
|\omega _{\pm }\rangle &=&|\omega \rangle +\frac 1{\omega \pm
io-H_0}V|\omega _{\pm }\rangle ,  \nonumber \\
\langle \omega _{\pm }| &=&\langle \omega |+\langle \omega _{\pm }|V\frac
1{\omega \mp io-H_0},
\end{eqnarray}
satisfying: 
\[
\langle \omega _{\pm }|\omega _{\pm }^{\prime }\rangle =\delta (\omega
-\omega ^{\prime }),\qquad |\omega _{+}\rangle =S(\omega )|\omega
_{-}\rangle ,\qquad \langle \omega _{+}|=S^{*}(\omega )\langle \omega _{-}| 
\]

In the last expressions, $S(\omega )$ is the trace of the S-matrix,
satisfying: 
\[
S(\omega )\,S^{*}(\omega )=1 
\]

If the Lipmann-Schwinger solutions form a complete orthonormal set we have,
in weak sense (as in eqs. 2.8 and 2.9 ): 
\[
I=\int_0^\infty d\omega |\omega _{\pm }\rangle \langle \omega _{\pm
}|,\qquad H=\int_0^\infty d\omega |\omega _{\pm }\rangle \omega \langle
\omega _{\pm }|. 
\]
These are the usual spectral expansions.

\subsection{Complex spectral decomposition of H.}

Let us now consider vectors $|\psi \rangle \in \Psi \subset {\cal S\subset H}
$ and $|\varphi \rangle \in \Phi \subset {\cal S\subset H}$, where $\Psi $
and $\Phi $ are defined in such a way that $\langle \omega _{\pm }|\psi
\rangle $ and $\langle \varphi |\omega _{\pm }\rangle $ have analytic
extensions $\langle z_{\pm }|\psi \rangle $ and $\langle \varphi |z_{\pm
}\rangle $ at least in the shadowed area of fig. 1 where the analytic
extension $S(z)$ of $S(\omega )$ has no poles. Repeating the formalism of
section 2.2, the functionals $|z_{\pm }\rangle $ and $\langle z_{\pm }|$ can
be used in the expansions of $|\psi \rangle $ and $\langle \varphi |$, that
is: 
\begin{equation}
|\psi \rangle =\int_\Gamma dz|z_{\pm }\rangle \langle z_{\pm }|\psi \rangle
,\qquad \langle \varphi |=\int_\Gamma dz\langle \varphi |z_{\pm }\rangle
\langle z_{\pm }|,
\end{equation}
from which we obtain: 
\begin{equation}
\langle \varphi |\psi \rangle =\int_\Gamma dz\langle \varphi |z_{\pm
}\rangle \langle z_{\pm }|\psi \rangle
\end{equation}
provided that: 
\begin{equation}
\langle z_{\pm }|z_{\pm }^{\prime }\rangle =\delta _\Gamma (z-z^{\prime }).
\end{equation}

We also have the generalized spectral decompositions: 
\begin{equation}
I=\int_\Gamma dz|z_{\pm }\rangle \langle z_{\pm }|,\qquad H=\int_\Gamma
dz|z_{\pm }\rangle z\langle z_{\pm }|,
\end{equation}
and the functionals $|z_{\pm }\rangle $ and $\langle z_{\pm }|$ are
generalized right and left eigenvectors of the total Hamiltonian $H$: 
\begin{eqnarray}
\langle z_{\pm }|H\,\psi \rangle &=&z\,\langle z_{\pm }|\psi \rangle \qquad
\psi \in \Psi  \nonumber \\
\langle \varphi |H\,z_{\pm }\rangle &\equiv &\langle H\,\varphi |z_{\pm
}\rangle =z\,\langle \varphi |z_{\pm }\rangle \qquad \varphi \in \Phi
\end{eqnarray}

We also have the relations: 
\begin{equation}
|z_{+}\rangle =S(z)|z_{-}\rangle ,\qquad \langle z_{+}|=S^{\#}(z)\langle
z_{-}|,
\end{equation}
being $S(z)$ and $S^{\#}(z)$ the meromorphic extensions of $S(\omega )$ and $%
S^{*}(\omega )$.

The previous equations are valid in a weak or functional sense and only in
the shadowed area of fig 1. In general the domains of analyticity of $%
\langle \varphi |z_{\pm }\rangle $ and $\langle z_{\pm }|\psi \rangle $
cannot be bigger than the domains in which $\langle \varphi |z\rangle $ and $%
\langle z|\psi \rangle $ are analytic, as can be seen from eq. 2.20, since
usually the second factors of the analytic extensions of the right hand
sides have poles. The domains of analyticity of $\langle \varphi |z_{\pm
}\rangle $ and $\langle z_{\pm }|\psi \rangle $ are a consequence of the
presence of these poles which are also poles of the S-matrix. For simplicity
we shall assume that $S(z)$ has a single pole at $z_0\in {\cal C}{\bf ^{-}}$
(the lower complex half plane) and therefore $S^{\#}(z)$ has a single pole
at $z_0^{*}\in {\cal C}{\bf ^{+}}$ (the upper complex half plane)$.$

Taking into account the presence of these poles different spectral
decompositions of $I$ and $H$ can be obtained. If we chose a curve $\Gamma
_d $, as in fig. $2_a$, the integral is equal to the sum of the
corresponding integrals over the curves $C_d$ and $C_d^{\prime }$ of fig. $%
2_b$ and we obtain a new spectral expansion for $I$ (see \cite{Bohm} for
more details): 
\begin{equation}
\begin{array}{c}
I=\int_{\Gamma _d}dz|z_{+}\rangle \langle z_{+}|=\int_{C_d}dz|z_{+}\rangle
\langle z_{+}|+\oint_{C_d^{\prime }}dzS(z)|z_{-}\rangle \langle z_{+}|= \\ 
=\int_{C_d}dz|f_z\rangle \langle \stackrel{\sim }{f_z}|+|f_0\rangle \langle 
\stackrel{\sim }{f_0}|,
\end{array}
\label{2.15}
\end{equation}
where 
\[
|f_z\rangle =|z_{+}\rangle ,\qquad \langle \stackrel{\sim }{f_z}|=\langle
z_{+}|, 
\]
\begin{equation}
|f_0\rangle =[-2\pi i(ResS)_{z_0}]^{\frac 12}|z_{0-}\rangle ,\qquad \langle 
\stackrel{\sim }{f_0}|=[-2\pi i(ResS)_{z_0}]^{\frac 12}\langle z_{0+}|.
\label{2.16}
\end{equation}

We emphasize that 2.27 is a formal expression which acquires meaning when it
is 'sandwiched' between $\langle \varphi |$ and $|\psi \rangle $ ($\varphi
\in \Phi $ and $\psi \in \Psi .$).

Orthogonality relations between the vectors defined in eqs. 2.28 can be
obtained using eqs.2.27 and 2.23. Consider, for example: 
\[
|f_0\rangle \langle \stackrel{\sim }{f_0}|f_0\rangle \langle \stackrel{\sim 
}{f_0}|=\oint_{C_d^{\prime }}dz\oint_{C_d^{\prime }}dz^{\prime
}|z_{+}\rangle \langle z_{+}|z_{+}^{\prime }\rangle \langle z_{+}^{\prime
}|= 
\]
\[
=\oint_{C_d^{\prime }}dz\oint_{C_d^{\prime }}dz^{\prime }\delta
_{C_d^{\prime }}(z-z^{\prime })|z_{+}\rangle \langle z_{+}^{\prime }|= 
\]
\[
=\oint_{C_d^{\prime }}dz|z_{+}\rangle \langle z_{+}|=|f_0\rangle \langle 
\stackrel{\sim }{f_0}| 
\]
From the first and last term of this equation we can deduce that: 
\begin{equation}
\stackrel{\sim }{\langle f_0}|f_0\rangle =1,  \label{2.17a}
\end{equation}
Using analogous arguments we can obtain: 
\[
\langle \stackrel{\sim }{f_0}|f_z\rangle =\langle \stackrel{\sim }{f_z}%
|f_0\rangle =0, 
\]
\begin{equation}
\langle \stackrel{\sim }{f_z}|f_{z^{\prime }}\rangle =\delta
_{C_d}(z-z^{\prime }).  \label{2.17b}
\end{equation}
In terms of the functionals defined in eq. 2.28 we obtain a new spectral
expansion for the Hamiltonian: 
\begin{equation}
H=z_0|f_0\rangle \langle \stackrel{\sim }{f_0}|+\int_{C_d}dz|f_z\rangle
z\langle \stackrel{\sim }{f_z}|,  \label{2.18}
\end{equation}
From this spectral expansion the time-evolution of any vector $|\psi \rangle
\in \Psi $ can be computed as: 
\begin{equation}
|\psi _t\rangle =e^{-iHt}|\psi \rangle =e^{-iz_0t}|f_0\rangle \langle 
\stackrel{\sim }{f_0}|\psi \rangle +\int_{C_d}dz|f_z\rangle e^{-izt}\langle 
\stackrel{\sim }{f_z}|\psi \rangle ,  \label{2.19}
\end{equation}
where we can identify an exponentially decaying component $|f_0\rangle ,$
that we would like to associate to an unstable state or Gamov vector.

We can also choose a curve $\Gamma _u$ as in fig. $3_a,$ and in this case we
obtain the contribution of the two curves $C_u$ and $C_u^{\prime }$ as shown
in fig. $3_b$. The new spectral decomposition so obtained reads: 
\[
I=\int_{\Gamma _u}dz|z_{+}\rangle \langle z_{+}|=\int_{C_u}dz|z_{+}\rangle
\langle z_{+}|+\int_{C_u^{\prime }}dz|z_{+}\rangle \langle z_{-}|S^{\#}(z)= 
\]
\begin{equation}
=\int_{C_u}dz|\stackrel{\sim }{f_z}\rangle \langle f_z|+|\stackrel{\sim }{f_0%
}\rangle \langle f_0|  \label{2.20}
\end{equation}
where: 
\[
|\stackrel{\sim }{f_z}\rangle =|z_{+}\rangle ,\qquad \qquad \qquad \langle
f_z|=\langle z_{+}| 
\]
\begin{equation}
|\stackrel{\sim }{f_0}\rangle =[2\pi i(ResS^{\#})z_0^{*}]^{\frac
12}|z_{0+}^{*}\rangle ,\qquad \langle f_0|=[2\pi i(ResS^{\#})z_0^{*}]^{\frac
12}\langle z_{0-}^{*}|,  \label{2.21}
\end{equation}
Using eqs. 2.33 and 2.23 we can also obtain: 
\[
\langle f_z|\stackrel{\sim }{f_{z^{\prime }}}\rangle =\delta
_{C_u}(z-z^{\prime }),\qquad \langle f_0|\stackrel{\sim }{f_0}\rangle =1, 
\]
\begin{equation}
\langle f_0|\stackrel{\sim }{f_{z^{\prime }}}\rangle =\langle f_z|\stackrel{%
\sim }{f_0}\rangle =0.  \label{2.22}
\end{equation}
In this representation the spectral expansion of the Hamiltonian is: 
\begin{equation}
H=z_0^{*}|\stackrel{\sim }{f_0}\rangle \langle f_0|+\int_{C_u}dz|\stackrel{%
\sim }{f_z}\rangle z\langle f_z|,  \label{2.23}
\end{equation}
and the evolution of a vector $|\varphi \rangle \in \Phi $ is given by: 
\begin{equation}
|\varphi _t\rangle =e^{-iHt}|\varphi \rangle =e^{-iz_0^{*}t}|\stackrel{\sim 
}{f_0}\rangle \langle f_0|\varphi \rangle +\int_{C_u}dze^{-izt}|\stackrel{%
\sim }{f_z}\rangle \langle f_z|\varphi \rangle ,  \label{2.24}
\end{equation}
In this case we find a growing component $|\stackrel{\sim }{f_0}\rangle $
that we shall identify also with an unstable state.

In this section $\Psi $ is the space of vectors $\psi $ for which $\langle
z_{+}|\psi \rangle $ is analytic in the region between $C_d$ and ${\cal R}%
^{+}$ of figure 2b, satisfying $e^{-iHt}\,\Psi \subset \Psi $, since if $%
\langle z_{+}|\psi \rangle \in \Psi $ $\langle z_{+}|e^{-iHt}\psi \rangle
=e^{-izt}\langle z_{+}|\psi \rangle $ and therefore the time evolution given
by 2.32 is valid for all values of $t\in {\cal R}$.

Also $\Phi $ is the set of vectors $\varphi $ for which $\langle
z_{+}|\varphi \rangle $ is analytic in the region between $C_u$ and ${\cal R}%
^{+}$ of figure 3b, $e^{-iHt}\,\Phi \subset \Phi $ and the time evolution
given by 2.37 is valid for all values of $t\in {\cal R}$.

If further restrictions are imposed on the vector spaces $\Psi $ and $\Phi $%
, the time evolutions 2.32 and 2.37 will be valid only for restricted values
of $t$. This possibility is discussed in sections 2.5 and 4.

The formal developments of this section are applied in section 5 to
Friedrichs model.

\subsection{Generalized expansions in the literature.}

The generalized spectral decompositions 2.31 and 2.36, obtained above,
appear in the literature originated by different approaches:

-Sudarshan et al. \cite{Sudar} proposed a generalized quantum formulation
using analytic continuations defined, from the beginning, on a curve in the
complex plane, like $\Gamma ,$ instead of the real semiaxis.

-A. Bohm et al. \cite{Bohm} considered a formulation of quantum mechanics in
rigged Hilbert spaces and fix $\Gamma =(-\infty ,0].$ In this approach
expressions 2.31 and 2.32 correspond to vectors $|\psi \rangle \in \phi
_{-}\subset \Psi \subset {\cal S\subset H}$, being $\phi _{-}$ the set of
vectors $|\psi \rangle $ such that $\langle \omega _{+}|\psi \rangle $ is a
function belonging to the Hardy class from below (cf. section 4). Then if 
$\vert\psi \rangle \in \phi _{-}$, $e^{-iHt}|\psi \rangle \in
\phi _{-}$ only for $t>0$. Expressions 2.36 and 2.37 apply to vectors $%
|\varphi \rangle \in \phi _{+}\subset \Phi \subset {\cal S\subset H}$, where 
$\phi _{+}$ is the set of vectors $|\varphi \rangle $ such that the function 
$\langle \omega _{+}|\varphi \rangle $ belongs to the Hardy class from
above. Then if $|\varphi \rangle \in \phi _{+}$, $e^{-iHt}|\varphi \rangle
\in \phi _{+}$ only for $t<0$. Thus the time evolution is decomposed in two
semigroups, being this fact the main advantage of Bhom proposal.

-The same representations are obtained by Petrosky et al. \cite{Petro}, for
the Friedrichs model, using a perturbative scheme together with a time
ordering rule. The interpretation of this approach in terms of rigged
Hilbert spaces is given by Antoniou et al. \cite{anto}.

-More recently, A. Bohm et al. \cite{Bohmanto} and A. Bohm \cite{referee}
deduced the need of Hardy class functions from a 'quantum arrow of time',
stating that measurements can only be realized after preparation of states.

-A more general mathematical structure (doublets) to represent Gamov vectors
is proposed in reference \cite{Dome}

The results presented in the previous subsection can be compared with the
ones of references \cite{anto} and \cite{Petro} if we define: 
\[
|f_\omega \rangle =\stackrel{\sim }{S(\omega )}|\omega _{-}\rangle ,\qquad
\langle \stackrel{\sim }{f_\omega }|=\langle \omega _{+}|, 
\]
\begin{equation}
|\stackrel{\sim }{f_\omega }\rangle =|\omega _{+}\rangle ,\qquad \langle
f_\omega |=\langle \omega _{-}|\stackrel{\sim }{S^{*}(\omega )}  \label{2.25}
\end{equation}
where $\stackrel{\sim }{S(\omega )}$ and $\stackrel{\sim }{S^{*}(\omega )}$
are the distributions defined by: 
\[
\int_0^\infty d\omega \stackrel{\sim }{S(\omega )}\psi (\omega
)=\int_0^\infty d\omega S(\omega )\psi (\omega )+2\pi i(ResS)_{z_0}\psi
(z_0), 
\]
\begin{equation}
\int_0^\infty d\omega \stackrel{\sim }{S^{*}(\omega )}\varphi (\omega
)=\int_0^\infty d\omega S^{*}(\omega )\varphi (\omega )-2\pi
i(ResS^{\#})_{z_0^{*}}\varphi (z_0^{*}),  \label{2.26}
\end{equation}
where $\psi (\omega )\in \Psi $ and $\varphi (\omega )\in \Phi .$

Using eqs. 2.38 and 2.39, we can rewrite eqs.2.31, 2.32, 2.36, and 2.37 as: 
\begin{equation}
H=z_0|f_0\rangle \langle \stackrel{\sim }{f_0}|+\int_0^\infty d\omega
|f_\omega \rangle \omega \langle \stackrel{\sim }{f_\omega }|,  \label{2.18'}
\end{equation}
\begin{equation}
e^{-iHt}|\psi \rangle =e^{-iz_0t}|f_0\rangle \langle \stackrel{\sim }{f_0}%
|\psi \rangle +\int_0^\infty d\omega e^{-i\omega t}|f_\omega \rangle \langle 
\stackrel{\sim }{f_\omega }|\psi \rangle ,  \label{2.19'}
\end{equation}
\begin{equation}
H=z_0^{*}|\stackrel{\sim }{f_0}\rangle \langle f_0|+\int_0^\infty d\omega |%
\stackrel{\sim }{f_\omega }\rangle \omega \langle f_\omega |,  \label{2.23'}
\end{equation}
\begin{equation}
e^{-iHt}|\varphi \rangle =e^{-iz_0^{*}t}|\stackrel{\sim }{f_0}\rangle
\langle f_0|\varphi \rangle +\int_0^\infty d\omega e^{-i\omega t}|\stackrel{%
\sim }{f_\omega }\rangle \langle f_\omega |\varphi \rangle .  \label{2.24'}
\end{equation}
Orthogonality conditions are: 
\[
\langle \stackrel{\sim }{f_0}|f_0\rangle =1,\qquad \langle \stackrel{\sim }{%
f_\omega }|f_{\omega ^{\prime }}\rangle =\delta (\omega -\omega ^{\prime }), 
\]
\[
\langle \stackrel{\sim }{f_0}|f_\omega \rangle =\langle \stackrel{\sim }{%
f_\omega }|f_0\rangle =0, 
\]
\[
\langle f_0|\stackrel{\sim }{f_0}\rangle =1,\qquad \langle f_\omega |%
\stackrel{\sim }{f_{\omega ^{\prime }}}\rangle =\delta (\omega -\omega
^{\prime }), 
\]
\begin{equation}
\langle f_0|\stackrel{\sim }{f_\omega }\rangle =\langle f_\omega |\stackrel{%
\sim }{f_0}\rangle =0.  \label{2.27}
\end{equation}

\section{The norm of Gamov vectors.}

Let us consider again eq. 2.32 and the decaying component $|f_0\rangle .$
The exponential decay is usually obtained in quantum mechanics as an
approximation given by the Fermi golden rule. However, for very short and
very large times, quantum mechanics predicts a deviation from exponential
behavior. As the life-time of some unstable states can be very large, and
the exponential decaying is measured with high precision, there has been
strong interest to consider generalized spectral decompositions of the
Hamiltonian, with complex eigenvalues, so that the corresponding
eigenvectors could describe unstable states with exact exponential decay,
namely the component $|f_0\rangle $ that we have found in eq 2.32. In order
to precise the nature of this kind of states it is interesting to compute
their norm and their mean energy.

Let us consider the ''Gamov vector'' $|f_0\rangle $ given by eq. 2.28, and
let us try to compute the norm and the energy of this state. The vectors $%
|f_0\rangle $ and $\langle f_0|$ defined in eqs. 2.28 and 2.34 can be
written as: 
\[
|f_0\rangle =[-2\pi i(ResS)_{z_0}]^{-\frac 12}\oint_{C_d^{\prime
}}|z_{+}\rangle dz, 
\]
\begin{equation}
\langle f_0|=[2\pi i(ResS^{\#})_{z_0^{*}}]^{-\frac 12}\oint_{C_u^{\prime
}}\langle z_{+}|dz.  \label{3.1}
\end{equation}
The integrals over $C_d^{\prime }$ and $C_u^{\prime }$ can be deformed into
a single closed curve $C,$ as shown in fig. 4\footnote{%
This deformation can be done provided spaces $\Phi $ and $\Psi $ are chosen
in such a way that the curve $C$ is contained in the domain where the test
functions $\langle \varphi |z\rangle $ ($\langle z|\psi \rangle $) of $%
|f_0\rangle $ ($\langle f_0|$) are analytic. It is also necesary that $%
|z_{+}\rangle $ and $\langle z_{+}|$ remain analytic during the deformation.
We shall check these requirements for Friedrichs model in section 5.}. Then
using eqs \ref{3.1} and 2.23 we obtain: 
\[
\langle f_0|f_0\rangle \sim \oint_{-C}dz\langle z_{.+}|\oint_Cdz^{\prime
}|z_{+}^{\prime }\rangle = 
\]
\begin{equation}
=-\oint_Cdz\oint_Cdz^{\prime }\delta _C(z-z^{\prime })=-\oint_Cdz=0.
\label{3.2}
\end{equation}
The same result can be obtained using the orthogonality condition 2.44: 
\[
\langle f_0|f_0\rangle =\langle f_0|I|f_0\rangle =\langle f_0|\int_0^\infty
d\omega |\omega _{+}\rangle \langle \omega _{+}|f_0\rangle = 
\]
\begin{equation}
=\int_0^\infty d\omega \langle f_0|\stackrel{\sim }{f_\omega }\rangle
\langle \stackrel{\sim }{f_\omega }|f_0\rangle =0,  \label{3.2'}
\end{equation}
and also: 
\begin{equation}
\langle f_0|H|f_0\rangle =\int_0^\infty d\omega \omega \langle f_0|\stackrel{%
\sim }{f_\omega }\rangle \langle \stackrel{\sim }{f_\omega }|f_0\rangle =0,
\label{3.3}
\end{equation}
We shall check these results again in section 5 in the case of the
Friedrichs model.

The fact that the generalized state $|f_0\rangle $ have zero norm and zero
energy seems to indicate that it is not a physical state, in fact:

i.-It is defined just as a functional and therefore it belongs to $\Psi
^{\times }$, a space of generalized states.

ii.-$|f_0\rangle $ never appears alone, but only as a component of a regular
physical state, as in eqs. 2.32 or 2.41. Thus $|f_0\rangle $ could be
considered as a ''ghost''. Like Fadeev-Popov ghosts, it is only useful to
perform certain calculation, precisely the generalized spectral
decompositions, in our case.

The functional $\langle \widetilde{f_0}|$ has the same properties.

Moreover these results are essential for the internal coherence of the
theory for two very important reasons:

i.-Eqs. 3.2 and 3.4 are necessary conditions for the conservation of
probability and energy within the theory. In fact the terms like $%
e^{-iHt}|f_0\rangle =e^{-iz_0t}|f_0\rangle $ vanish when $t\rightarrow
\infty ,$ therefore they must have vanishing norm and energy, since these
quantities must be constant in time.

ii.-For mixed states, eqs. 3.2 and 3.4 appear again. Traces and mean values
of ''fluctuations'' vanish, and this fact is also essential for the
consistency of the formalism, as we shall show in \cite{Laura}.

\section{Time evolution and time asymmetry.}

A. Bohm et al., in reference \cite{Bohm}, \cite{Bohmanto} and \cite{referee}
proposed a special choice for the spaces we have called $\Phi $ and $\Psi .$
Precisely they proposed that these spaces must coincide with spaces $\phi
_{+}$ and $\phi _{-}$ defined as the sets of vectors with analytic extension
to the upper or the lower half plane, or in mathematical terms: 
\begin{equation}
\phi _{\pm }=\{|\psi \rangle /\langle \omega _{\mp }|\psi \rangle \in \theta
({\cal S}\cap H_{\pm }^2)\},  \label{4.1}
\end{equation}
where ${\cal S}$ denotes the Schwarz class, $H_{\pm }^2$ the upper (lower)
Hardy class, and $\theta $ is the Heaviside step function.

A complex function $f(\omega )$ on ${\cal R}$ is a Hardy class function from
above (below) if:

i.-$f(\omega )$ is the boundary value of a function $f(z)$ of complex
variable $z=x+iy$ that is analytic in the half plane $y>0$ $(y<0).$

ii.-$\int_{-\infty }^{+\infty }|f(x+iy)|^2dx<k<\infty ,$ for all $y$ such
that $0<y<\infty ,$ $(-\infty <y<0).$

Then, $|\psi \rangle \in \phi _{-}\Rightarrow e^{-iHt}|\psi \rangle \in \phi
_{-}$ if $t>0,$ ( \cite{Bohm},\cite{anto}) and the time evolution of $|\psi
\rangle $ can be computed using eqs. 2.32 or 2.41 where $|f_0\rangle ,$ $%
\langle \stackrel{\sim }{f_0}|,$ $|f_\omega \rangle $, and $\langle 
\stackrel{\sim }{f_\omega }|$ are generalized states of the dual space $\phi
_{-}^{\times }$. Thus: 
\begin{equation}
e^{-iHt}\phi _{-}\subset \phi _{-},\quad {\rm if }t>0,  \label{4.2}
\end{equation}
Also $|\varphi \rangle \in \phi _{+}\Rightarrow e^{-iHt}|\varphi \rangle \in
\phi _{+}$ if $t<0$, (\cite{Bohm},\cite{anto}) and the time evolution of $%
|\varphi \rangle $ can be computed using eqs. 2.37 or 2.43, being $|%
\stackrel{\sim }{f_0}\rangle ,$ $\langle f_0|,$ $|\stackrel{\sim }{f_\omega }%
\rangle ,$ and $\langle f_\omega |$ generalized states of the dual space $%
\phi _{+}^{\times }.$ Thus: 
\begin{equation}
e^{-iHt}\phi _{+}\subset \phi _{+},\quad {\rm if }t<0.  \label{4.3}
\end{equation}
In the first case the Gamov vector $|f_0\rangle $ decays toward the future,
while in the second case the Gamov vector $|\stackrel{\sim }{f_0}\rangle $
decays towards the past.

The demonstration of this properties, given in reference \cite{Bohm}, is
based in two theorems:

-{\bf Paley-Wiener theorem.}

If $\varphi $$(\omega )\in H_{-}^2$ (the Hardy class from below) then the
Fourier transform: 
\[
\lbrack {\cal F}\varphi (\omega )]_s=(2\pi )^{-\frac 12}\int_{-\infty
}^{+\infty }\exp (-is\omega )\varphi (\omega )d\omega 
\]
is endowed with the property: 
\[
\lbrack {\cal F}\varphi (\omega )]_s=0,\quad {\rm if }s>0. 
\]

-{\bf Theorem:}

If $\varphi (\omega )\in H_{-}^2$ then $\exp (-i\omega t)\varphi (\omega
)\in H_{-}^2$ if $t>0.$

In fact, if $\varphi (\omega )\in H_{-}^2$then: 
\[
\lbrack {\cal F}\exp (-i\omega t)\varphi (\omega )]_s=(2\pi )^{-\frac
12}\int_{-\infty }^{+\infty }\exp [-i\omega (s+t)]\varphi (\omega )d\omega 
\]
which is zero if $s>-t$ and consequently $\exp (-i\omega t)\varphi (\omega
)\in H_{-}^2$ if $t\rangle 0.$ Q.E.D.

As both $\phi _{+}$ and $\phi _{-}$are dense in ${\cal H}$, it seems
reasonable to represent any physical state as a vector of $\phi _{+}$ or $%
\phi _{-},$ so we will restrict all our physical reasonings to one of these
spaces only. In fact, it is quite useless to speculate about the physical
nature of spaces $\phi _{+}^{\times }$ or $\phi _{-}^{\times },$ since all
the physic is really contained, as we shall see, in one of the two test
functions spaces $\phi _{+}$ or $\phi _{-}$ and furthermore these spaces are
dense in $\phi _{+}^{\times }$ and $\phi _{-}^{\times }$.

Let us now analyze the action of the time inversion operator $K$ in the
spaces $\phi _{+}$ or $\phi _{-}$. For any $|\psi \rangle $ represented in
terms of the $|\omega \rangle $ ($H_0|\omega \rangle =\omega |\omega \rangle
)$ the time-inversion operator $K$ is defined by: 
\begin{equation}
K|\psi \rangle =\int_0^\infty d\omega |\omega \rangle \langle \omega |\psi
\rangle ^{*},  \label{4.4}
\end{equation}
from which we can deduce that $K|\omega \rangle =|\omega \rangle .$

The operator $K$ satisfies: 
\[
K(a_1|\psi _1\rangle +a_2|\psi _2\rangle )=a_1^{*}K|\psi _1\rangle
+a_2^{*}K|\psi _2\rangle 
\]
\[
\langle \varphi |K\psi \rangle =\langle K\varphi |\psi \rangle ^{*},\qquad
\qquad \qquad K^2=1. 
\]
If, as usual, $[H,K]=0$ (see eq. 1.2) and $|\psi (t)\rangle $ is a solution
of Schroedinger equation, $K|\psi (-t)\rangle $ is also a solution of this
equation, and the quantum model turns out to be time symmetric if we work in 
${\cal H}$ space, and we have that: 
\begin{equation}
K:{\cal H\rightarrow H},  \label{4.5}
\end{equation}
namely $K$ maps the space of what in a non rigorous theory are considered
physical solutions over the same space (and the same thing happens with $%
{\cal S}$ in the rigorous theory, cf. eq \ref{1.4b}). As we shall see in a
moment, this is not the case if the space of the physical states is $\phi
_{-}$or $\phi _{+}.$

Assuming $[H_0,K]=[V,K]=0,$ together with \ref{4.4} and the definition of $%
|\omega _{\mp }\rangle $ given in 2.20, we obtain: 
\[
K|\omega _{\pm }\rangle =|\omega _{\mp }\rangle 
\]
Let $|\psi \rangle \in \phi _{-}$, then $\langle \omega _{+}|\psi \rangle
\in \theta ({\cal S\cap }H_{-}^2),$ and 
\[
K|\psi \rangle =K\int_0^\infty d\omega |\omega _{+}\rangle \langle \omega
_{+}|\psi \rangle =\int_0^\infty d\omega |\omega _{-}\rangle \langle \omega
_{+}|\psi \rangle ^{*} 
\]
from which $\langle \omega _{-}|K\psi \rangle =\langle \omega _{+}|\psi
\rangle ^{*}\in \theta ({\cal S}\cap H_{+}^2),$ and therefore $K|\psi
\rangle \in \phi _{+}.$ In general we have proved that: 
\begin{equation}
K:\phi _{\pm }\rightarrow \phi _{\mp }.  \label{4.6}
\end{equation}
We have now all the elements to show that if we postulate that: e. g. $\phi
_{-}$ is the space of the physical states our theory becomes
time-asymmetric, as was stated in the introduction. In fact:

i.-From eq. \ref{4.6} we see that $K:\phi _{-}\rightarrow \phi _{+}$ so time
inversion do not exists within the space of physical states

ii.-If $t>0,$ eq. \ref{4.2} shows that the evolution operator $U(t)=e^{-iHt}$
exists in the physical space $\phi _{-}.$ But eq. \ref{4.3} shows also that
the inverted time operator $U(t)^{-1}=e^{iHt}$ do not exists in this space
of physical state $\phi _{-}.$ This is, of course, the essence of an
irreversible theory.

Then, with a minimal change we have obtain an irreversible quantum theory
and, as in the physical real world it is impossible to invert the time
evolution, we can claim that $\phi _{-}$ mimic, better than ${\cal H,}$ the
physical reality.

We close this section with three observations:

i.-Our model can be a local model, as a decaying (or growing) process, or a
global model, as a cosmological model of the universe.

-If it is a local model we will always deal with a period of time $-\infty
<t\leq 0$ when the system is prepared and $\phi _{+}$is chosen as the
physical space, and a period $0\leq t<\infty $ when the measurement is
performed and the physical space is $\phi _{-}$ \cite{Bohmanto} and \cite
{referee}. Of course in this case we cannot consider the problem of the
arrow of time in full, since we can change the roles of $\phi _{+}$ and $%
\phi _{-}$ and also we cannot consider e.g. the cosmological arrow of time.
Being our model just local, in this case, the real arrow of time is
essentially imposed from the exterior of the model, i. e. the rest of the
universe.

-In the second case we will consider only the period $0\leq t<\infty $ and $%
t=0$ will usually corresponds to the ''big bang''time, and we postulate that
the universe do not exists before that time. In this case the model is
complete but, of course, much more complicated, and the research of this
kind of problems is just beginning (\cite{Cosmo}). The space of physical
states is usually $\phi _{-}$ (even if we can also use $\phi _{+}$ and the
period $-\infty <t\leq 0,$ as we shall explain in ii). Therefore
irreversibility is introduced in the model as we have explained and the
problem of the arrows of time in the universe can be explained in full. Once
the global arrow of time is established in the universe it can be used to
define the local arrows in the subsystems of the universe (using e. g. the
Reichenbach branch system \cite{Reichen}).

ii.-Someone may say that we have introduced in the global case the arrow of
time ''just by hand'', when we chose space $\phi _{-}$ or $\phi _{+}$ as the
space of physical states. In order to answer this criticism we must define
two important words ''conventional'' and ''substantial''. Precisely:

-In mathematics we use to work with identical objects, like points, the two
directions of an axis, the two semicones of a light cone etc.

-In physics there are also identical objects, like identical particles, the
two spin directions etc.

-When (\cite{Penrose},\cite{Sachs}) we are forced to call by different names
two identical objects we will say that we are establishing a {\it %
conventional difference}, while

- if we call by different names two different objects we will say that we
are establishing a {\it substantial difference.}

The problem of time asymmetry is that, in all time-symmetric physical
theories, usually the difference between past and future is just
conventional. In fact, we can change the word ''past'' by the word
''future'', in these theories, and nothing changes. But we have the clear
psychological filling that the past is substantially different than the
future. Thus the problem of the arrow of time is to find theories where past
is substantially different than future and such that usual well established
physics remains valid. Our minimal irreversible quantum mechanics is one of
these theories.

In fact: the difference between $\phi _{-}$ and $\phi _{+}$ in the global
case is just conventional since these two spaces are identical. Thus physics
is the same in $\phi _{-}$than in $\phi _{+}.$ Think in a cosmological
model, life will be the same in the universe of $\phi _{-}$ than in the
universe of $\phi _{+}.$ In fact, since in both models of universe (if
completely computed) all the arrows of time must point in the same
direction, there is no physical way to decide if we are in one model or the
other. Thus the choice between $\phi _{-}$ and $\phi _{+}$ is just
conventional and physically irrelevant. But once this choice is made a
substantial difference is established in the model of the universe e. g. the
only time evolution operator is $U(t)=e^{-iHt},t>0,$ and cannot be inverted,
we have equilibrium only towards the future, etc.

Thus the choice between $\phi _{-}$ and $\phi _{+}$ is trivial and
unimportant in the global case, that is why the arrow of time is not
introduced by hand. The important choice is between ${\cal H}$ (or ${\cal S}$%
) and $\phi _{-}$ (or $\phi _{+})$ as the space of our physical states. And
we are free to make this choice, since a good physical theory begins by the
choice of the best mathematical structure to mimic real nature.

iii.- Eq. \ref{3.2} is valid if the curve $C$ is contained in a domain where
the functions of both spaces $\Psi $ and $\Phi $ are analytic. But if $\Psi
=\phi _{-}$ and $\Phi =\phi _{+},$ since the domain where $\phi _{-}$ is
analytic do not overlap with the one of $\phi _{+},$ it is impossible to
draw the curve $C$. Nevertheless if we define the domain of analyticity of
the functions of $\Psi $ and $\Phi $ in such a way that they overlap and
contains the curve $C$ and if the domain of $\Psi $ ($\Phi $) contains the
lower (upper) half plane, then: 
\[
\Psi \subset \phi _{-}\subset {\cal H}\subset \phi _{-}^{\times }\subset
\Psi ^{\times } 
\]
\[
\Phi \subset \phi _{+}\subset {\cal H}\subset \phi _{+}^{\times }\subset
\Phi ^{\times } 
\]
Therefore eq. \ref{3.2} is valid also if we only consider that $|f_0\rangle
\in \phi _{-}^{\times }$ since this space is contained in $\Psi ^{\times }$
where eq. \ref{3.2} was demonstrated. Analogously we can prove that $|%
\widetilde{f_0}\rangle $ has also vanishing norm.

\section{Friedrichs model.}

In section 3 we have shown that our theory is feasible, in the sense that
the norm and the probability is conserved, and in section 4 we have
demonstrated that we are dealing with a quantum irreversible theory. It is
now necessary to prove that the physical results, obtained by this theory,
are correct. This task have been partially done already and can be found in
the literature \cite{Sudar},\cite{Bohm},\cite{anto}, where it is
demonstrated that these results coincide, in general, with the ones of the
usual ''coarse-graining'' theory (\cite{20},\cite{Paz}). In fact, the main
contribution of this paper is only to give a better theoretical foundation
to these results. So we close this paper with just one example, the well
known Friedrichs model, namely the simplest model of an unstable state
coupled with a continuous ''radiation'' field. We shall show how our theory
leads to the decaying of the unstable state of this model and how, what is
left of it, reappears in the continuous spectrum, we would say as radiation.

Let us consider the Hamiltonian: 
\begin{equation}
H=m|1\rangle \langle 1|+\int_0^\infty d\omega |\omega \rangle \omega \langle
\omega |+\int_0^\infty d\omega V_\omega (|\omega \rangle \langle
1|+|1\rangle \langle \omega |),  \label{5.1}
\end{equation}
where $|1\rangle $ is the unstable state, $\{|\omega \rangle \}$ $(0\leq
\omega <\infty )$ can be consider as the states of a continuous set of
oscillators that symbolizes a radiation field, and the last term of the r.
h. s. of eq. \ref{5.1} is the interaction term.

For this model Lipmann-Schwinger equations 2.20 can be solved exactly
giving: 
\begin{eqnarray}
|\omega _{\pm }\rangle &=&|\omega \rangle +\frac{V_\omega }{\eta (\omega \pm
i0)}\left[ |1\rangle -\int_0^\infty \frac{d\omega ^{\prime }V_{\omega
^{\prime }}|\omega ^{\prime }\rangle }{\omega ^{\prime }-(\omega \pm io)}%
\right] ,  \nonumber \\
\langle \omega _{\pm }| &=&\langle \omega |+\frac{V_\omega }{\eta (\omega
\mp i0)}\left[ \langle 1|-\int_0^\infty \frac{d\omega ^{\prime }V_{\omega
^{\prime }}\langle \omega ^{\prime }|}{\omega ^{\prime }-(\omega \mp io)}%
\right] ,
\end{eqnarray}
where: 
\begin{equation}
\eta (z)=z-m+\int_0^\infty \frac{d\omega ^{\prime }V_{\omega ^{\prime }}^2}{%
\omega ^{\prime }-z},\qquad z\in {\cal C-R}^{+}  \label{5.3}
\end{equation}

If we consider the set of vectors $\psi \in {\cal C}\times {\cal S}$ such
that $\langle 1|\psi \rangle \in {\cal C}$ and $\langle \omega |\psi \rangle
\in {\cal S}$, the vectors $|\omega _{\pm }\rangle $ ($\langle \omega _{\pm
}|$) are antilinear (linear) functionals acting on ${\cal C}\times {\cal S}$%
, and generalized right (left) eigenvectors of the Hamiltonian 5.1: 
\begin{equation}
H|\omega _{\pm }\rangle =\omega |\omega _{\pm }\rangle ,\qquad \langle
\omega _{\pm }|H=\omega \langle \omega _{\pm }|
\end{equation}

The vectors $|\omega _{+}\rangle $ and $|\omega _{-}\rangle $ are related by 
\[
|\omega _{+}\rangle =S(\omega )|\omega _{-}\rangle ,\qquad \qquad \qquad
S(\omega )=\frac{\eta (\omega -i0)}{\eta (\omega +i0)} 
\]

If $\eta (z)$ do not vanish for real values of $z$, it is possible to prove 
\cite{Sudar} that they form a complete biorthogonal set, in the sense that
for any two vectors $\varphi $ and $\psi $ in ${\cal C}\times {\cal S}$
results 
\[
\langle \varphi |\psi \rangle =\langle \varphi |1\rangle \langle 1|\psi
\rangle +\int d\omega \,\langle \varphi |\omega \rangle \langle \omega |\psi
\rangle =\int d\omega \,\langle \varphi |\omega _{\pm }\rangle \langle
\omega _{\pm }|\psi \rangle , 
\]
and therefore 
\[
I=\int d\omega \,|\omega _{\pm }\rangle \langle \omega _{\pm }|,\qquad
\langle \omega _{\pm }|\omega _{\pm }^{\prime }\rangle =\delta (\omega
-\omega ^{\prime }). 
\]

As $\eta (z)$ defined by 5.3 has a cut in ${\cal R}^{+}$, it is possible to
define the extension $\eta _{+}(z)$ ($\eta _{-}(z)$) from the upper to the
lower (lower to the upper) half plane as 
\begin{eqnarray}
\eta _{+}(z) &=&\left\{ 
\begin{array}{l}
\eta (z)\quad {\rm if }{\rm Im}\,z>0 \\ 
\eta (z)+2\pi iV_z^2\quad {\rm if }{\rm Im}\,z<0
\end{array}
\right\}  \nonumber \\
\eta _{-}(z) &=&\left\{ 
\begin{array}{l}
\eta (z)-2\pi iV_z^2\quad {\rm if }{\rm Im}\,z>0 \\ 
\eta (z)\quad {\rm if }{\rm Im}\,z<0
\end{array}
\right\}
\end{eqnarray}

We assume that $\eta _{+}(z)=0$ has a single solution $z_0\in {\cal C}^{-}$,
and therefore $\eta _{-}(z_0^{*})=0$, $z_0^{*}\in {\cal C}^{+}$. This means
that the analytic extension $S(z)=\frac{\eta _{-}(z)}{\eta _{+}(z)}$ of $%
S(\omega )=\frac{\eta (\omega -i0)}{\eta (\omega +i0)}$ is analytic in $%
{\cal C}-z_0$ with a simple pole at $z_0\in {\cal C}^{-}$.

The functionals 5.2 can be analytically extended to the complex plane: 
\begin{eqnarray}
|z^{+}\rangle &=&|z\rangle +\frac{V_z}{\eta _{+}(z)}\left[ |1\rangle
-\int_0^\infty d\omega ^{\prime }\,V_{\omega ^{\prime }}|\omega ^{\prime
}\rangle \left( \frac 1{\omega ^{\prime }-s}\right) _z^{+}\right] \\
\langle z^{+}| &=&\langle z|+\frac{V_z}{\eta _{-}(z)}\left[ \langle
1|-\int_0^\infty d\omega ^{\prime }\,V_{\omega ^{\prime }}\langle \omega
^{\prime }|\left( \frac 1{\omega ^{\prime }-s}\right) _z^{-}\right]
\end{eqnarray}

In the last expressions we introduced the functionals $\left( \frac 1{\omega
^{\prime }-s}\right) _z^{+}$ and $\left( \frac 1{\omega ^{\prime }-s}\right)
_z^{-}$ defined by: 
\begin{eqnarray}
\int_0^\infty d\omega ^{\prime }\left( \frac 1{\omega ^{\prime }-s}\right)
_z^{+}\varphi (\omega ^{\prime }) &=&\left\{ 
\begin{array}{l}
\int_0^\infty d\omega ^{\prime }\frac 1{\omega ^{\prime }-z}\varphi (\omega
^{\prime })\quad {\rm if }{\rm Im}z>0 \\ 
\int_0^\infty d\omega ^{\prime }\frac 1{\omega ^{\prime }-z}\varphi (\omega
^{\prime })+2\pi i\varphi (z)\quad {\rm if }{\rm Im}z<0
\end{array}
\right\}  \nonumber \\
\int_0^\infty d\omega ^{\prime }\left( \frac 1{\omega ^{\prime }-s}\right)
_z^{-}\psi (\omega ^{\prime }) &=&\left\{ 
\begin{array}{l}
\int_0^\infty d\omega ^{\prime }\frac 1{\omega ^{\prime }-z}\psi (\omega
^{\prime })-2\pi i\psi (z)\quad {\rm if }{\rm Im}z>0 \\ 
\int_0^\infty d\omega ^{\prime }\frac 1{\omega ^{\prime }-z}\psi (\omega
^{\prime })\quad {\rm if }{\rm Im}z<0
\end{array}
\right\}
\end{eqnarray}
which are well defined on functions $\varphi (\omega ^{\prime })$ ($\psi
(\omega ^{\prime })$), $\omega ^{\prime }\in {\cal R}^{+}$, having analytic
extensions to the lower (upper) half plane.

Following the procedure given in section 2, the spectral decompositions 2.31
and 2.36 can be found and the following generalized eigenvectors are
obtained (\cite{anto},\cite{Petro}): 
\[
|\stackrel{\sim }{f_0}\rangle =\frac 1{\sqrt{\eta _{-}^{\prime }(z_0^{*})}%
}\left[ |1\rangle -\int_0^\infty d\omega V_\omega \left( \frac 1{\omega
-s}\right) _{z_0^{*}}^{-}|\omega \rangle \right] 
\]
\[
\langle f_0|=\frac 1{\sqrt{\eta _{-}^{\prime }(z_0^{*})}}\left[ \langle
1|-\int_0^\infty d\omega V_\omega \left( \frac 1{\omega -s}\right)
_{z_0^{*}}^{-}\langle \omega |\right] 
\]
\[
|\stackrel{\sim }{f_\omega }\rangle =|\omega \rangle +\frac{V_\omega }{\eta
_{+}(\omega )}\left[ |1\rangle +\int_0^\infty \frac{d\omega ^{\prime
}V\omega ^{\prime }}{\omega -\omega ^{\prime }+io}|\omega ^{\prime }\rangle
\right] 
\]
\begin{equation}
\langle f_\omega |=\langle \omega |+\frac{V_\omega }{\widetilde{\eta
_{-}(\omega )}}\left[ \langle 1|+\int_0^\infty \frac{d\omega ^{\prime
}V_{\omega ^{\prime }}}{\omega -\omega ^{\prime }-io}\langle \omega ^{\prime
}|\right] ,  \label{5.5}
\end{equation}
\[
|f_0\rangle =\frac 1{\sqrt{\eta _{+}^{\prime }(z_0)}}\left[ |1\rangle
-\int_0^\infty d\omega V_\omega \left( \frac 1{\omega -s}\right)
_{z_0}^{+}|\omega \rangle \right] 
\]
\[
\langle \stackrel{\sim }{f_0}|=\frac 1{\sqrt{\eta _{+}^{\prime }(z_0)}%
}\left[ \langle 1|-\int_0^\infty d\omega V_\omega \left( \frac 1{\omega
-s}\right) _{z_0}^{+}\langle \omega |\right] 
\]
\[
|f_\omega \rangle =|\omega \rangle +\frac{V_\omega }{\widetilde{\eta
_{+}(\omega )}}\left[ |1\rangle +\int_0^\infty \frac{d\omega ^{\prime
}V\omega ^{\prime }}{\omega -\omega ^{\prime }+io}|\omega ^{\prime }\rangle
\right] 
\]
\begin{equation}
\langle \stackrel{\sim }{f_\omega }|=\langle \omega |+\frac{V_\omega }{\eta
_{-}(\omega )}\left[ \langle 1|+\int_0^\infty \frac{d\omega ^{\prime
}V\omega ^{\prime }}{\omega -\omega ^{\prime }-io}\langle \omega ^{\prime
}|\right] ,  \label{5.6}
\end{equation}

In the previous equations we introduced the distributions $\frac 1{%
\widetilde{\eta _{-}(\omega )}}$ and $\frac 1{\widetilde{\eta _{+}(\omega )}%
} $, defined by: 
\[
\int_0^\infty d\omega \frac 1{\widetilde{\eta _{+}(\omega )}}\varphi (\omega
)\equiv \int_0^\infty d\omega \frac 1{\eta _{+}(\omega )}\varphi (\omega
)+2\pi i\frac{\varphi (z_0)}{\eta _{+}^{\prime }(z_0)}, 
\]

\begin{equation}
\int_0^\infty d\omega \frac 1{\widetilde{\eta _{-}(\omega )}}\psi (\omega
)\equiv \int_0^\infty d\omega \frac 1{\eta _{-}(\omega )}\psi (\omega )-2\pi
i\frac{\psi (z_0^{*})}{\eta _{-}^{\prime }(z_0^{*})}.  \label{5.7d}
\end{equation}

From the explicit expressions given in 5.9 and 5.10 for $\langle f_0|$ and $%
|f_0\rangle $, and the orthogonality relations 
\[
\langle 1|1\rangle =1,\quad \langle 1|\omega \rangle =\langle \omega
|1\rangle =0,\quad \langle \omega |\omega ^{\prime }\rangle =\delta (\omega
-\omega ^{\prime }), 
\]
we obtain 
\begin{equation}
\langle f_0|f_0\rangle =\frac 1{\sqrt{\eta _{-}^{\prime }(z_0^{*})\eta
_{+}^{\prime }(z_0)}}\left[ 1+\int_0^\infty d\omega V_\omega ^2\left( \frac
1{\omega -s}\right) _{z_0^{*}}^{-}\left( \frac 1{\omega -s^{\prime }}\right)
_{z_0}^{+}\right]
\end{equation}

Taking into account 
\[
\frac 1{\omega -s}\times \frac 1{\omega -s^{\prime }}=\frac 1{s-s^{\prime
}}\left( \frac 1{\omega -s}-\frac 1{\omega -s^{\prime }}\right) , 
\]
and 
\[
\eta _{+}(z_0)=z_0-m+\int_0^\infty d\omega V_\omega ^2\left( \frac 1{\omega
-s^{\prime }}\right) _{z_o}^{+}=0, 
\]
\[
\eta _{-}(z_0^{*})=z_0^{*}-m+\int_0^\infty d\omega V_\omega ^2\left( \frac
1{\omega -s}\right) _{z_0^{*}}^{-}=0, 
\]
in equation 5.12, we obtain 
\[
\langle f_0|f_0\rangle =0. 
\]

Then, even if it was already proved in section 3, we show again, with this
example, that Gamov vector $|f_0\rangle $ has zero norm.

The spaces $\phi _{+}$ ($\phi _{-})$ in which the analytic extensions to the
upper (lower) half plane are well defined in this model are: 
\[
\phi _{\pm }=\{|\psi \rangle /\langle 1|\psi \rangle \in {\cal C}{\bf ,}%
\langle \omega |\psi \rangle \in \theta ({\cal S}\cap H_{\pm }^2)\}. 
\]
It is interesting to note that a state with $\langle \omega |\psi \rangle =0$
can be considered to be either in $\phi _{-}$ or in $\phi _{+},$ because the
analytic extension of this last function is $\langle z|\psi \rangle =0.$
Consequently $|1\rangle \in \phi _{-}\cap \phi _{+},$ and $e^{-iHt}|1\rangle 
$ is in $\phi _{-}$ for $t>0$ and in $\phi _{+}$ for $t<0,$ therefore it
decays to the future as a vector of $\phi _{-},$ and to the past, as a
vector of $\phi _{+}.$

Let us now compute the time evolution of the observables $A$ and their mean
values. We have: 
\begin{equation}
\langle A\rangle _t=\langle \psi (t)|A|\psi (t)\rangle =\langle \psi
|e^{iHt}Ae^{-iHt}|\psi \rangle =\langle \psi |A(t)|\psi \rangle .
\label{5.8}
\end{equation}

We assume that the observable $A$ can be written as: 
\[
A=A_1|1\rangle \langle 1|+\int_0^\infty d\omega A_\omega |\omega \rangle
\langle \omega |+\int_0^\infty d\omega A_{1\omega }|1\rangle \langle \omega
|+ 
\]
\begin{equation}
+\int_0^\infty d\omega ^{\prime }A_{\omega ^{\prime }1}|\omega ^{\prime
}\rangle \langle 1|+\int_0^\infty \int_0^\infty d\omega d\omega ^{\prime
}A_{\omega ^{\prime }\omega }|\omega ^{\prime }\rangle \langle \omega |,
\label{5.9}
\end{equation}
where, in the second term, we have included a singular diagonal term, which
is present in many observables, like the Hamiltonian of eq. \ref{5.1}.
Alternatively we may use the vectors $|\omega _{+}\rangle =|\stackrel{\sim }{%
f_\omega }\rangle $ and $\langle \omega _{+}|=\langle \stackrel{\sim }{%
f_\omega }|$ to represent $A$ as: 
\begin{equation}
A=\int_0^\infty d\omega A_\omega ^{+}|\stackrel{\sim }{f_\omega }\rangle
\langle \stackrel{\sim }{f_\omega }|+\int_0^\infty \int_0^\infty d\omega
d\omega ^{\prime }A_{\omega \omega ^{\prime }}^{+}|\widetilde{f_\omega }%
\rangle \langle \widetilde{f_\omega }|=A_I+A_F.  \label{5.10}
\end{equation}
Comparing eq. \ref{5.10} with eq. \ref{5.9}, and using the explicit
expressions for $|\widetilde{f_\omega }\rangle $ and $\langle \widetilde{%
f_\omega }|$ given by eqs. 5.9 and 5.10 we can prove that $A_\omega
^{+}=A_\omega ,$ and therefore: 
\begin{equation}
\int_0^\infty d\omega \Pi _\omega A=A_I=\int_0^\infty A_\omega |\widetilde{%
f_\omega }\rangle \langle \widetilde{f_\omega }|,  \label{5.11}
\end{equation}
where: 
\[
\Pi _\omega A\equiv A_\omega |\widetilde{f_\omega }\rangle \langle 
\widetilde{f_\omega }| 
\]
This is the time invariant part of the observable $A$ in the Heisemberg
picture. Using eqs. 5.15 and 5.4 we obtain: 
\begin{equation}
A(t)=e^{iHt}Ae^{-iHt}=A_I+e^{iHt}A_Fe^{-iHt}.  \label{5.12}
\end{equation}
Essentially observables are represented by operators used to compute mean
values, as in eq. 5.13. If in this equation $\psi \in $$\phi _{-}$ and we
wish to use the spectral decomposition 2.41, it is necessary to assume that
the functions $A_{1\omega },$ $A_{\omega ^{\prime }1,}$ and $A_{\omega
^{\prime }\omega }$ of eq. 5.14 can be analytically extended to the lower
(upper) half plane in the variable $\omega $ ($\omega ^{\prime }).$(we shall
discuss this fact at large in a forthcoming paper \cite{Laura}). This
assumption is compatible with the results of reference \cite{Bohmanto} and 
\cite{referee}.So we can deduce that: 
\begin{eqnarray}
&&e^{iHt}A_Fe^{-iHt}  \nonumber \\
&=&e^{iHt}\left[ |\widetilde{f_0}\rangle \langle f_0|+\int d\omega |%
\widetilde{f_\omega }\rangle \langle f_\omega |\right] A_F\left[ |f_0\rangle
\langle \widetilde{f_0}|+\int d\omega ^{\prime }|f_{\omega ^{\prime
}}\rangle \langle \widetilde{f_{\omega ^{\prime }}|}\right] e^{-iHt} 
\nonumber \\
&=&e^{i(z_0^{*}-z_0)t}\Pi _{00}A+\int_0^\infty d\omega e^{i(\omega
-z_0)t}\Pi _{\omega 0}A+  \nonumber \\
&&+\int_0^\infty d\omega ^{\prime }e^{i(z_0^{*}-\omega ^{\prime })t}\Pi
_{0\omega ^{\prime }}A+\int_0^\infty \int_0^\infty e^{i(\omega -\omega
^{\prime })t}\Pi _{\omega \omega ^{\prime }}A,
\end{eqnarray}

where: 
\[
\Pi _{00}A\equiv |\widetilde{f_0}\rangle \langle f_0|A_F|f_0\rangle \langle 
\widetilde{f_0}|, 
\]
\[
\Pi _{\omega 0}A\equiv |\widetilde{f_\omega }\rangle \langle f_\omega
|A_F|f_0\rangle \langle \widetilde{f_0}|, 
\]
\[
\Pi _{0\omega }A\equiv |\widetilde{f_0}\rangle \langle f_0|A_F|f_\omega
\rangle \langle \widetilde{f_\omega }|, 
\]
\begin{equation}
\Pi _{\omega \omega ^{\prime }}A\equiv |\widetilde{f_\omega }\rangle \langle
f_\omega |A_F|f_{\omega ^{\prime }}\rangle \langle \widetilde{f_{\omega
^{\prime }}}|.  \label{5.14}
\end{equation}
From eqs. 5.17 and 5.18 we can deduce the time evolution of an observable,
and from eq. 5.13 the time evolution of its mean value.

But in order to understand what really is going on it is interesting to
obtain an approximated expression for $A(t),$ when the interaction function $%
V_\omega $ is small, because in this case the unstable state $|1\rangle $
will have a long life-time and we will be able to obtain a long pure
exponential decay. For this purpose it is necessary to obtain the asymptotic
form of the projectors of eqs. 5.16 and 5.19 for weak interactions. Taking
into account eq. 5.9 and 5.10 we obtain the weak limit: 
\[
\lim_{V\rightarrow 0} |\widetilde{f_\omega }\rangle \langle 
\widetilde{f_\omega }|=|\omega \rangle \langle \omega |+\lim_{%
V\rightarrow 0} \frac{V_\omega ^2}{\eta _{+}(\omega )\eta _{-}(\omega )%
}|1\rangle \langle 1| 
\]
Then, from eq (5.3) we have: 
\[
\frac{V_\omega ^2}{\eta _{+}(\omega )\eta _{-}(\omega )}=\frac{V_\omega ^2}{%
(\omega -m-\Delta )(\omega -m-\Delta ^{*})}=\frac{V_\omega ^2}{\Delta
^{*}-\Delta }\left( \frac 1{\omega -m-\Delta ^{*}}-\frac 1{\omega -m-\Delta
}\right) 
\]
where: 
\[
\Delta \equiv \int_0^\infty \frac{d\omega ^{\prime }V_{\omega ^{\prime }}^2}{%
\omega +io-\omega ^{\prime }}. 
\]
Therefore: 
\[
\Delta ^{*}-\Delta =\int_0^\infty d\omega ^{\prime }V_{\omega ^{\prime
}}^2\left( \frac 1{\omega -\omega ^{\prime }-io}-\frac 1{\omega -\omega
^{\prime }+io}\right) =2\pi iV_\omega ^2 
\]
As $Im\Delta ^{\star }\rangle 0$ and $\lim_{V\rightarrow 0}
\Delta ^{*}=io^{+},$ we obtain: 
\[
\lim_{V\rightarrow 0} \frac{V_\omega ^2}{\eta _{+}(\omega )\eta
_{-}(\omega )}=\frac 1{2\pi i}\left( \frac 1{\omega -m-io}-\frac 1{\omega
-m+io}\right) =\delta (\omega -m) 
\]
and therefore: 
\begin{equation}
\lim_{V\rightarrow 0} |\widetilde{f_\omega }\rangle \langle 
\widetilde{f_\omega }|=|\omega \rangle \langle \omega |+\delta (\omega
-m)|1\rangle \langle 1|.  \label{5.15}
\end{equation}
From eqs. 5.9 and 5.10 we have: 
\begin{equation}
\lim_{V\rightarrow 0} |\widetilde{f_0}\rangle \langle f_0|=%
\lim_{V\rightarrow 0} |f_0\rangle \langle \widetilde{f_0}%
|=|1\rangle \langle 1|,  \label{5.16}
\end{equation}
\begin{equation}
\lim_{V\rightarrow 0} |\widetilde{f_\omega }\rangle \langle
f_\omega |=\lim_{V\rightarrow 0} |f_\omega \rangle \langle 
\widetilde{f_\omega }|=|\omega \rangle \langle \omega |,  \label{5.17}
\end{equation}
These results can be used to obtain the limits of the projectors of eqs.
5.16 and 5.19: 
\[
\lim_{V\rightarrow 0} \Pi _\omega A=A_\omega [|\omega \rangle
\langle \omega |+\delta (\omega -m)|1\rangle \langle 1|] 
\]
\[
\lim_{V\rightarrow 0} \Pi _{00}A=(A_1-A_{\omega =m})|1\rangle
\langle 1|, 
\]
\[
\lim_{V\rightarrow 0} \Pi _{0\omega }A=A_{1\omega }|1\rangle
\langle \omega |, 
\]
\[
\lim_{V\rightarrow 0} \Pi _{\omega 0}A=A_{\omega 1}|\omega
\rangle \langle 1|, 
\]
\begin{equation}
\lim_{V\rightarrow 0} \Pi _{\omega \omega ^{\prime }}A=A_{\omega
\omega ^{\prime }}|\omega \rangle \langle \omega ^{\prime }|.  \label{5.18}
\end{equation}
When the interaction vanishes $z_0\rightarrow m.$ However this would be a
bad choice for the values in eq. 5.18 if we want to know the approximate
behavior of $A(t)$ for $t\rightarrow \infty .$ Precisely, if we solve: 
\[
\eta _{+}(z_0)=z_0-m-\int_0^\infty d\omega V_\omega ^2\left( \frac
1{s-\omega }\right) _{z_0}^{+}=0 
\]
up to the second order, we obtain: 
\begin{equation}
z_0\cong m+\int_0^\infty \frac{d\omega V_\omega ^2}{m+io-\omega },\qquad
z_0^{*}-z_0\cong 2\pi iV_m^2,  \label{5.19}
\end{equation}
Replacing the results of eqs. 5.23 and 5.24 in eqs. 5.18 and 5.16 we obtain
the approximate behavior of $A(t)$ when $V\rightarrow 0,$ and $t\rightarrow
\infty ,$ in such a way that $V_m^2t$ is finite$.$ E.g.: if we choose $%
A=|1\rangle \langle 1|$ or $A=|\omega \rangle \langle \omega |$ (or in a
more rigorous way $A=\int f(\omega )|\omega \rangle \langle \omega |d\omega
),$ we can compute the probability to find the system in the unstable states 
$|1\rangle $ or in the continuous ''radiation'' field $|\omega \rangle $ for
large $t$ and small $V,$ precisely: 
\[
\langle \psi (t)|1\rangle \langle 1|\psi (t)\rangle \cong e^{-2\pi
V_m^2t}\langle \psi |1\rangle \langle 1|\psi \rangle 
\]
\begin{equation}
\langle \psi (t)|\omega \rangle \langle \omega |\psi (t)\rangle \cong
\langle \psi |\omega \rangle \langle \omega |\psi \rangle +(1-e^{-2\pi
V_m^2t})\delta (\omega -m)\langle \psi |1\rangle \langle 1|\psi \rangle .
\label{5.20}
\end{equation}
These equations clearly show the exponential decay of the unstable state $%
|1\rangle $ and the simultaneous appearance of a radiation state at the
energy $\omega =m$, namely the radiation outcome of the unstable state. This
is a completely reasonable and experimentally verified physical result.
However, we must observe that in this case the pure exponential behavior is
only a consequence of the approximation we have used.

\section{Conclusion.}

Peter Bergmann said in 1967(\cite{Mackey}):

{\it ''It is not very difficult to show that the combination of the
reversible laws of mechanics with Gibbsian statistics does not lead to
irreversibility but that the notion of irreversibility must be added as an
special ingredient...}

...{\it the explanation of irreversibility in nature is to my mind still
open''}

In fact, from a reversible classical or quantum theory it is impossible to
obtain an irreversible one, making only mathematical manipulation. The
theory will remain always reversible. So necessarily a {\it new ingredient }%
must be added. From 1967 these ingredients were found and classified:
coarse-graining, traces, stochastic noises, etc. \cite{Mackey},\cite{Lasota}%
. The problem is to know what is the minimal ingredient that produces
irreversibility in the more aesthetic and economical way. We propose that
this minimal ingredient, is to change space ${\cal S}$, satisfying \ref{1.4b}%
, by space $\phi _{-}$, satisfying \ref{1.5}. This modification can be done
in the microscopical quantum level and it is simply to change the space of
the regular physical states, as we have done in this paper. The deep
physical meaning of this change is the following:

Experimentally we can only perform a finite number of measurements, so that,
when we (indirectly) measure a quantum state we only know a finite number of
points (or data) of the corresponding wave function. As we want to have the
whole wave function, because, e. g.,we cannot find the derivative of a set
of finite points, we interpolate this set with a function endowed with
mathematical properties that we can freely chose according to our
conveniences. We can choose this function in space ${\cal S}$ , namely to
use the Gel'fand triplet \ref{1.1}, and then we will obtain the usual
reversible quantum mechanics. In this case, if we want to take into account
irreversible processes using the usual formalism, we are forced to
coarse-grain the system. But we can directly interpolate using functions of
the space $\phi _{-}$, namely using the Gel'fand triplet \ref{1.6}, and we
will obtain an irreversible quantum mechanics, from the very beginning.
Clearly the second process is more economical than the first one.

Thus, the physical basis of the two approaches is the same, we always have
only a finite amount of information, but the way to deal with this fact is
different. (We will farther discuss these matters elsewhere.) As far as we
know the two approaches yield the same physical results, since up to now we
do not know of a ''cross experiment'' to tell us which formalism is the good
one. So both theories seems physically equivalent. But, even if the first
theory perhaps is more intuitive, the second one have two advantages:

i.-It contains just one fundamental modification, as we have explained.

ii.-It provide us with a very simple and powerful computational method: the
spectral decomposition in Gamov vectors and the corresponding time evolution
(eqs. 2.32, 2.37), obtained using just analytic continuation, which is much
more easy to handle than e.g. Feynman path integral, used in the first
theory.

So we believe that now the reader know almost all the features of the
problem and he can reach to a final decision by himself.

\section{Acknowledgment.}

This work was partially supported by grants: CI1-CT94-0004 of the European
Community, PID-0150 of CONICET (National Research Council of Argentina),
EX-198 of Buenos Aires University, and 12217/1 of Fundaci\'{o}n Antorchas.

\section{Figure caption.}

fig.1: The curve $\Gamma .$

fig. $2_a$: The curve $\Gamma _d$.

fig. $2_b$. The curves $C_d$ and $C_d^{\prime }$.

fig. $3_a.$ The curve $\Gamma _u$.

fig. $3_b$. The curves $C_u$ and $C_u^{\prime }$.

fig. 4. The curve $C.$


\section{Bibliography.}

\begin{thebibliography}{99}
\bibitem{Bogo}  Bogolubov N. N., Logunov A. A., Todorov I.T. {\it %
Introduction to axiomatic quantum field theory,} Benjamin, London, (1975).

\bibitem{20}  Zwanzig R. W. J. Chem. Phys., {\bf 33}, 1338, (1960).

Zwanzig R. W. {\it Statistical mechanics of irreversibility}, in Quantum
statistical mechanics, Meijer P.- ed., Gordon and Breach, New York, (1966).

Zeh D. H. {\it The physical basis of the direction of time,}
Springer-Verlag, Berlin, (1989).

Zurek W., Physics Today, Oct. 1991, 36.

Halliwell J. J. et al. ed., {\it Physical origin of time asymmetry,}
Cambridge Univ. Press, Cambridge, (1994).

\bibitem{Paz} Hu B.L., Paz J.P., Zhang Y., Phys. Rev. D {\bf 45}, 2843 (1992).

\bibitem{Sudar}  Sudarshan E. C. G., Chiu C. B., Gorini V., Phys. Rev. D, 
{\bf 18}, 2914, (1978).

Parravicini G., Gorini V., Sudarshan E. C. G., J. Math. Phys. {\bf 21},
2208, (1980).

Sudarshan E.C.G., Chiu C. B.,{\it \ Analytic continuation of quantum systems
and their temporal evolution,} Preprint DOE-402000-276 CPP/91/15, Univ. of
Texas, Austin, 1992. Phys Rev. D {\bf 47,} 2602, (1993).

\bibitem{Bohm}  Bohm A., {\it Quantum mechanics: foundations and
applications,} Springer-Verlag, Berlin, (1986).

Bohm A., Gadella M., Maynland B. G., Am. J. Phys. {\bf 57}, 1103, (1989).

Gadella M., J. Math. Phys., {\bf 22,} 1462, (1981).

Gadella M., J. Math. Phys., {\bf 24,} 2124, (1983).

Gadella M., J. Math. Phys., {\bf 25,} 2461, (1984).

\bibitem{anto}  Antoniou I., Prigogine I., Physica A, {\bf 192}, 443, (1993).

Antoniou I., Suchanecki Z, Laura R., Tasaki S.,{\it Quantum systems with
diagonal singularity,} U. L. B. Preprint, (1995).

\bibitem{Laura}  Laura R. Castagnino M, Minimal Irreversible Quantum
Mechanics: Mixed States (in preparation)

\bibitem{Petro}  Petrosky T. Y., Prigogine I., Tasaki S., Physica A, {\bf 173%
}, 175, (1991).

\bibitem{Bohmanto}  Bohm A., Antoniu I., Kielanowsski P., J. Math. Phys. 
{\bf 36,} 1, (1995).

\bibitem{referee}  Bohm A., Phys. Rev. A, {\bf 51}, 1758, (1995).

\bibitem{Dome}  Castagnino M.,\ Domenech G., Levinas M.L., Umerez N., Jour.
Math. Phys. {\bf 37}, 2107 (1996).

\bibitem{Cosmo}  Castagnino M. Gunzig E., Nardone P., Prigogine I, Tasaki S.,%
{\it \ Quantum cosmology and large Poincar\'{e} systems, }in Quantum, Chaos
and Cosmology, Namiki M., ed., AIP book division, New York, (1996).

Castagnino M., Gunzig E., Lombardo F., Gen. Rel and Grav, {\bf 27,} 257,
(1995).

Castagnino M. Lombardo F., Gen. Rel. and Grav. {\bf 28},263, (1996).

Castagnino M., Gaioli F., Gunzig E, Found. of Cos. Phys., {\bf 16}, 221,
(1995).

Castagnino M., Laura R., {\it The cosmological essence of time asymmetry},
Proc. SILARG\ VIII, Ed. Rodrigues W., World Scientific, Singapore, (1993).

\bibitem{Reichen}  Reichenbach H., ${\em The}${\em \ direction of time},
Univ. of California Press (1956).

Davies P., Stirring up trouble, in {\em 'Physical origin of time asymmetry'}%
, Halliwell J.\ J. et al. eds., Cambridge Univ. Press (1996).

Castagnino M., {\em The global nature of the arrow of time and
Bohm-Reichenbach diagram.} Submitted to the Proceedings of XXI Int.
Colloquium on Group Theoretical Methods in Physics, Goslar (1996).

\bibitem{Penrose}  Penrose R.,{\it Singularities and time asymmetry, } in
General relativity, an Einstein centenary survey, Hawking S., Israel W. ed.,

\bibitem{Sachs}  Sachs R. G., {\it The physics of time reversal, }Univ. of
Chicago press, Chicago, (1987).

\bibitem{Mackey}  Mackey M. C., {\it Time's arrow: the origins of
thermodynamic behavior, }Springer-Verlag, Berlin, (1992).

\bibitem{Lasota}  Lasota A. Mackey M. C., {\it Probabilistic properties of
deterministic systems, }Cambridge Univ. press, Cambridge, (1985).
\end{thebibliography}
\end{document}